\documentclass{optica-article}

\journal{opticajournal} 

\articletype{Research Article}

\usepackage{color}
\usepackage{amsmath}
\usepackage{graphicx}
\usepackage{booktabs}
\usepackage{multirow}
\usepackage{amsfonts}
 \usepackage{soul}

\begin{document}

\title{High resolution up-conversion imaging in the $10\ \mu m$ band under incoherent illumination}

\author{Zhao-Qi-Zhi Han,\authormark{1,2,3,\dag} Xiao-Hua Wang,\authormark{1,2,3,\dag} Jin-Peng Li,\authormark{1,2,3} Bo-Wen Liu,\authormark{1,2,3} Zheng-He Zhou,\authormark{1,2,3} He Zhang,\authormark{1,2,3} Yin-Hai Li,\authormark{1,2,3,4} Zhi-Yuan Zhou,\authormark{1,2,3,4,*} and Bao-Sen Shi\authormark{1,2,3,5}}

\address{\authormark{1}Laboratory of Quantum Information, University of Science and Technology of China, Hefei 230026, China\\
\authormark{2}CAS Center for Excellence in Quantum Information and Quantum Physics, University of Science and Technology of China, Hefei 230026, China\\
\authormark{3}Anhui Province Key Laboratory of Quantum Network, University of Science and Technology of China, Hefei 230026, China\\
\authormark{4}Anhui Kunteng Quantum Technology Co. Ltd., Hefei 231115, China\\
\authormark{5}drshi@ustc.edu.cn\\
\authormark{\dag}These authors contribute equally to this work.\\

\email{\authormark{*}zyzhouphy@ustc.edu.cn}} 


\begin{abstract*} 

Long-wavelength infrared band exhibits significant utility in thermal signature acquisition and molecular spectral analysis, among other applications. The up-conversion detection technique enables effective signal transduction into the detection bandwidth of silicon-based photodetectors, thereby facilitating high-sensitivity photonic measurements. We realized high-resolution up-conversion imaging for incoherent thermal targets in the $10\ \mu m$ spectral regime for the first time. Furthermore, this work presents the first derivation of analytical models characterizing depth of field and astigmatic aberration in up-conversion imaging systems, which show excellent agreement between theoretical and experimental results. The results demonstrate generalisability to various up-conversion imaging systems, thus providing critical insights for the design and optimisation of such systems.

\end{abstract*}

\section{Introduction}
The mid-infrared (MIR) region (typically referring to wavelengths of $2.5 - 25 \mu m$) has emerged as a key area for multi-domain technological applications because of its distinctive properties. It encompasses dense molecular vibrations and rotational energy levels, exhibiting enhanced absorption and emission characteristics in comparison to other bands \cite{rasskazov2017ac, ronen2015mt}. These spectral lines have been shown to be capable of uniquely identifying various chemical bonds, thus making them a valuable asset in a range of fields, including molecular recognition \cite{ropcke2006pssat, alicja2022sabc, du2019as} and biomedical research \cite{an2023asr,razumtcev2022ac}. This band covers the peak area of black-body radiation from objects at room temperature (approximately $300\ K$), rendering it an ideal band for thermal radiation detection. This feature can be employed to facilitate non-contact temperature monitoring, including real-time tracking of various disasters (including natural disasters \cite{hua2017jfr,ouzounov2004asr} and engineering disasters \cite{xu2025tust}) and environmental changes \cite{pejcic2009sensor, ali2023ao}. The location of the band is in the transparent window of atmosphere and this renders it a widely used component in remote sensing applications \cite{walsh2016jl,thomson1993rse}.

In order to further improve the application capacity of the MIR band, it is essential to employ high-performance detectors in the aforementioned precision measurement scenarios. However, traditional semiconductor infrared photon detectors based on photoelectric effects, such as mercury cadmium telluride (MCT), currently face a series of problems. These include low detection efficiency, high noise, long response time and so on. And as the wavelength of the signal increases, the performance of these materials deteriorates significantly, and their detection performance in the long-wavelength infrared (LWIR, generally referring to $8 - 14 \mu m$) region is limited. The type-\uppercase\expandafter{\romannumeral2} superlattice (T2SL) exhibits superior performance in the LWIR spectrum when compared to MCT \cite{alshahrani2021apr}. However, T2SL remains encumbered by challenges such as intricate growth processes, elevated costs, a paucity of pixels and so on \cite{ai2024afm}. Consequently, the up-conversion detection (UCD) scheme emerges as a viable supplementary solution. The UCD scheme utilises the sum-frequency generation (SFG) process to up-convert MIR signals into the visible band while preserving its original information \cite{han2023adi},  and high-performance silicon-based detectors are then employed for detection. This technology has been widely applied in mid-wavelength infrared (MWIR, generally referring to $2.5 - 8 \mu m$) detection. Efficient single-mode up-conversion detection has been achieved in the waveguide system \cite{neely2012ol, buchter2009nonlinear}. As for bulk crystals, imaging with a large field of view and high sensitivity has been achieved by combining different pump types such as ultrafast pulses \cite{zhou2013apl, huang2022nc,zheng2024optica, zhang2025lpr} and continuous wave \cite{israelsen2019lsa} pump. In the MWIR region, a substantial theoretical and experimental works has been reported by using UCD for coherent and incoherent radiation targets \cite{han2024prd, ge2023prapplied, dam2012np, dam2012oe}. In the quantum field, this technology has also helped people achieve the detection of photon pairs within this band \cite{mancinelli2017nc, ge2024sa,li2024quantum, cai2022pr}. The UCD of LWIR is a subject that is currently being investigated by many groups. Current research predominantly centers on optimizing single-pixel detection performance \cite{tidemand2024ol, john2021lpr, tidemand2024inverstigation, guo2023ipt}. As for LWIR up-conversion imaging, the high-resolution imaging by focused coherent beam illuminated has been achieved \cite{tidemand2024long}. However, to our best knowledge, there is no report on up-conversion imaging of incoherent thermal radiation targets in the LWIR region.

This work achieved high-resolution up-conversion imaging of incoherent thermal radiation targets in the LWIR band and systematically investigated key resolution-limiting factors, including numerical aperture, depth of field, and astigmatism for the first time. In this paper, the theoretical limit of spatial resolution in our UCD system was achieved under incoherent illumination in $10 \ \mu m$ band. It has been demonstrated that the resolution has almost reached its theoretical limits, and can be further improved by optimising the collection lens. Furthermore, the impact of the shape transformation of pump beam on system resolution was investigated, it was discovered that the transformation of the pump beam shape has the capacity to induce alterations in system resolution in different directions. We measured, for the first time, the ratio between the average intensity at the effective aperture boundary and the peak intensity of the Gaussian pump beam, a critical parameter governing the waist design. The present study constitutes the inaugural investigation into the impact of aberrations on the resolution of up-conversion imaging systems. Quantitative calculation equations for depth of field and astigmatism have been provided for the first time, and the experimental results are in good agreement with theory. We also studied other parameters in up-conversion imaging of incoherent radiation targets. This work will provide significant theoretical and experimental support for the design and practical application of conversion systems in the future.

\section{Theory and Simulation}

In our system, up-conversion imaging is based on the SFG process that occurs in AGS crystals. Here, mid-infrared signal photons with a frequency of $\omega_{MIR}$ are converted into sum-frequency photons with a frequency of $\omega_{SF}$ under the action of a strong pump beam with a frequency of $\omega_{p}$, for which the principle of energy conservation is strictly satisfied, i.e. $\omega_{SF} = \omega_{p} + \omega_{MIR}$. The sum-frequency photons can be detected by high-performance Silicon-based scientific complementary metal-oxide semiconductor (sCMOS) camera to achieve room temperature and high sensitivity detection.

\begin{figure*}
  \centering\includegraphics[width=\textwidth]{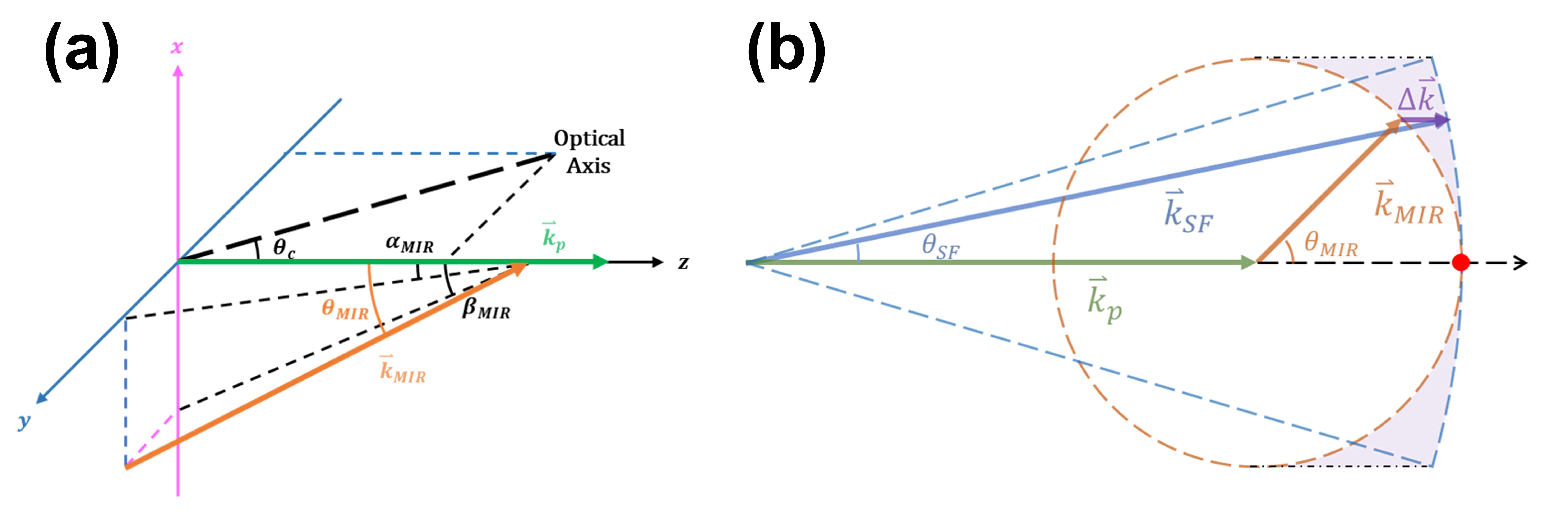}
  \caption{Schematic diagram of non-collinear phase match. (a) Schematic diagram of propagation direction of mid-infrared signal and pump beam in AGS crystal. (b) Phase match between the MIR signal beam and pump beam in the plane in which they are located when the collinear phase matching condition is precisely met (Red dots represent phase matching points).}
  \label{fig:boat1}
\end{figure*}

 In the imaging process, mid-infrared signal photons propagate in multiple directions, so phase match must be considered under non-collinear conditions. \textbf{Figure \ref{fig:boat1}} shows the schematic diagrams of non-collinear phase match.
As shown in \textbf{Figure \ref{fig:boat1}(a)}, the optical axis is located in the $yz$ plane at an angle $\theta_c$ to the $z$-axis; the pump beam propagates along the $z$-axis. The angle between the mid-infrared signal beam and the $z$-axis is $\theta_{MIR}$, while the angles between its projections on the $yz$ and $xz$ planes and the $z$-axis are $\alpha_{MIR}$ and $\beta_{MIR}$, respectively. The component perpendicular to the z-axis direction of $\vec{k}_{SF}$ only comes from $\vec{k}_{MIR}$\cite{ge2023prapplied}, so the phase match during the up-conversion process can be expressed as:
\begin{equation}
\left\{ \begin{gathered}
  {k_{SF}}\sin {\theta _{SF}} = {k_{MIR}}\sin {\theta _{MIR}} \hfill \\
  {k_{SF}}\cos {\theta _{SF}} + \Delta k = {k_{MIR}}\cos {\theta _{MIR}} + {k_p} \hfill \\ 
\end{gathered}  \right.
\end{equation}
Of these, $\vec{k}_{i} \left( i=MIR,\ p,\ SF \right)$ represents the wave vectors of the mid-infrared signal beam, the pump beam and the sum-frequency beam, respectively. And ${k_i} = \frac{{2\pi {n_i}}}{{{\lambda _i}}}\left( i=MIR,\ p,\ SF \right)$, where ${n_i}$ represents their refractive index in the propagation direction and $\lambda_i$ represents their wavelength. It is important to note that, given the optical axis is not perpendicular to the propagation direction, the refractive index of extraordinary light varies with the propagation angle. For more detailed information, please refer to the supplementary materials. $\Delta k$ represents phase mismatch during SFG process. $\theta_{SF}$ is the angle between the sum-frequency beam and the $z$-axis. The aforementioned parameters are delineated in \textbf{Figure \ref{fig:boat1}(b)}. The orange and blue dashed lines in \textbf{Figure \ref{fig:boat1}(b)} represent the possible directions of $\vec{k}_{MIR}$ and $\vec{k}_{SF}$, respectively. The purple shaded area represents the size and range of $\Delta k$ as the angle increases.

\begin{figure*}
  \centering\includegraphics[width=\textwidth]{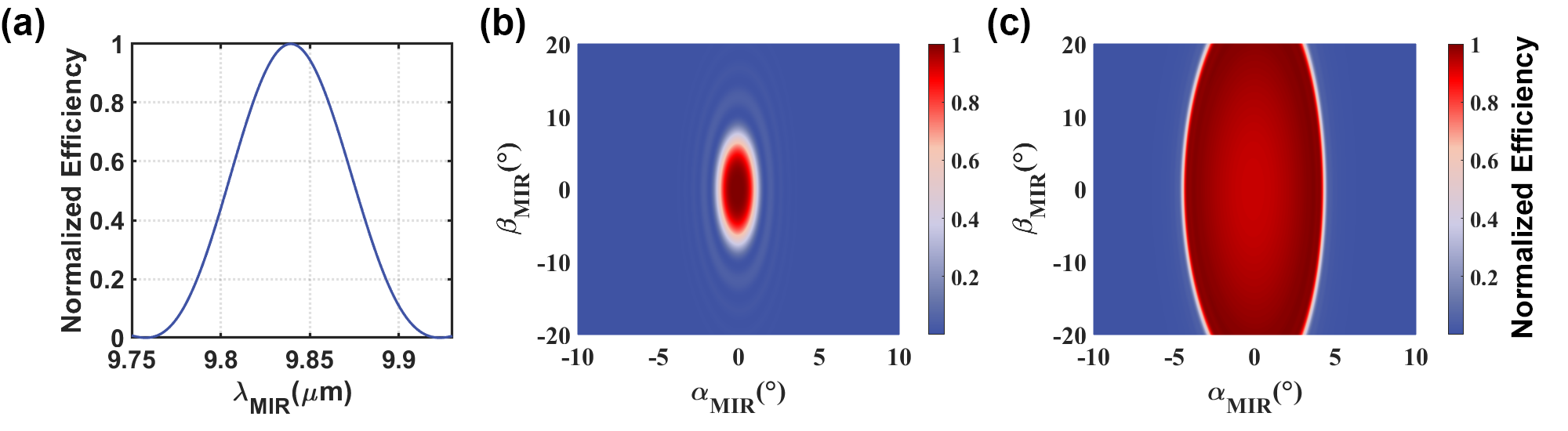}
  \caption{Numerical calculation results of wavelength and angle conversion bandwidth. (a) The relationship between normalized conversion efficiency and the wavelength of MIR signal when its incident angle is $\theta=0$. (b) The relationship between conversion efficiency and the incident angle when the wavelength of the CW coherent MIR signal precisely satisfies the collinear phase-matching condition. (c) The relationship between conversion efficiency and the incident light angle under incoherent thermal radiation illumination (The converted wavelength bandwidth of MIR signal is $600\ nm$).}
  \label{fig:boat2}
\end{figure*}

The conversion efficiency of SFG process is proportional to $\text{sinc} ^2\left( {\frac{{\Delta kL}}{2}} \right)$\cite{boyd2020nonlinear, shen1984principles}. Pursuant to the above analysis, the wavelength bandwidth and angle bandwidth can be calculated under different conditions\cite{tidemand2016josab, takaoka1999ao}. The simulation results are displayed in \textbf{Figure \ref{fig:boat2}}. All theoretical calculations are based on the cutting angle $\theta=40.5^\circ,\ \varphi=0^\circ$ of the type-\uppercase\expandafter{\romannumeral2} AGS crystal. Theoretical calculations show that, compared to continuous-wave (CW) coherent illumination case based on AGS crystals\cite{tidemand2024aplp}, using incoherent thermal radiation illumination has significant advantages in terms of wavelength and angle conversion bandwidth. As shown in \textbf{Figure \ref{fig:boat2}(b-c)}, for instance, the full width of half-maximum (FWHM) of conversion angle is $2.50^\circ$ in $\alpha_{MIR}$ direction under CW coherent illumination, and it is $8.92^\circ$ in the case under incoherent thermal radiation illumination with wavelength bandwidth of $600 \ nm$ in calculation. For more information on the center wavelength, wavelength bandwidth, and field of view of this UCD process, please refer to the supplementary materials.

\section{Experimental Setup and Results}

\subsection{Experimental Setup}
\begin{figure*}
  \centering\includegraphics[width=\textwidth]{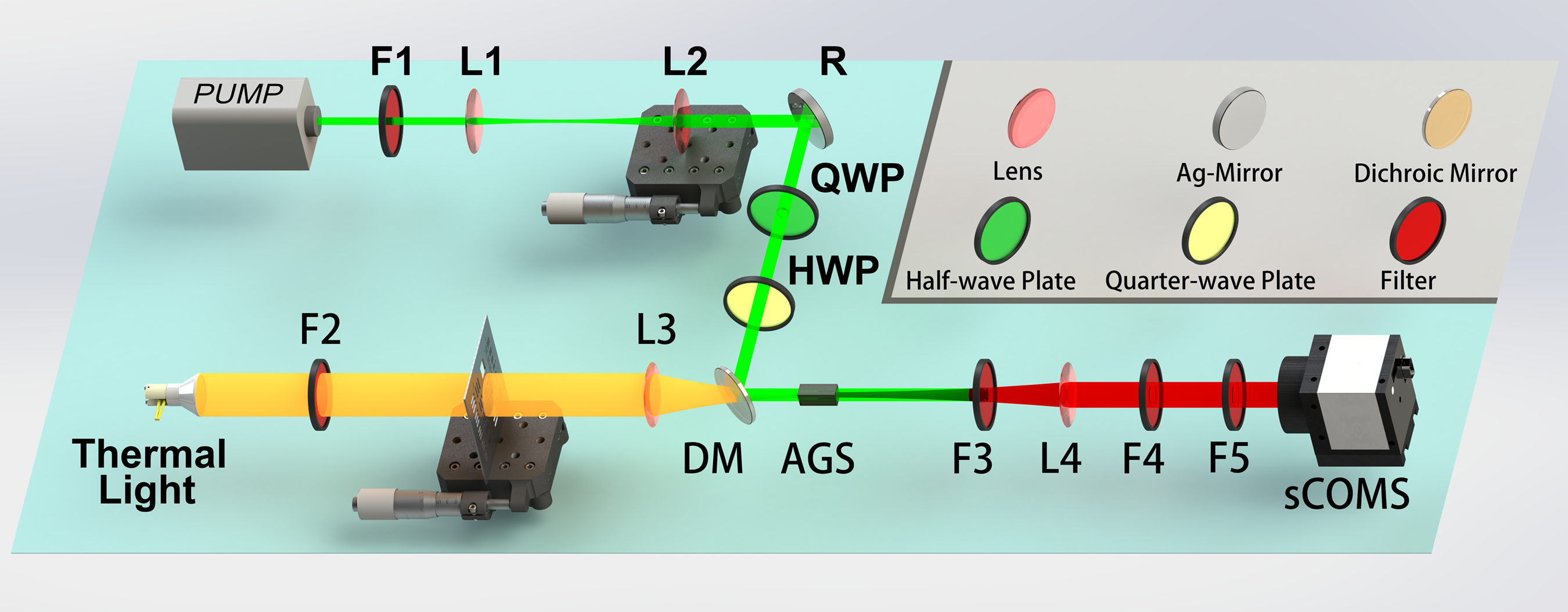}
  \caption{Schematic diagram of the experimental setup. L terms: the lenses; DM: dichromatic mirror; F: filters; R: mirror; HWP: half-wave plate; QWP: quarter-wave plate; AGS: $AgGaS_2$ crystal.}
  \label{fig:boat3}
\end{figure*}

The schematic diagram of the experimental setup is shown in \textbf{Figure \ref{fig:boat3}}. The pump is a CW source generated by injecting seed light from a semiconductor laser (KT-TSL-1064, Anhui Kunteng Quantum Technology) into a Yb-doped fibre amplifier. F1 is a long pass filter with the cut-on wavelength of $1000\ nm$ (FELH1000, Thorlabs). And L2 is placed on a displacement table. The MIR source is generated by a thermal infrared emitter (HIS2000R-0, Infrasolid). A long-pass filter with cut-on wavelength of $6000\ nm$ (Stock \#36-151, Edmund) labelled F2 is used to obtain the MIR signal within the conversion bandwidth range. L3 and L4 form a $4f$ system. Their common focus is located at the centre of the AGS crystal. The imaging object and sCMOS camera (Dhyana95 V2, Tucsen) are placed at suitable planes, respectively. The imaging target is also placed on a displacement table. F3 and F4 are short-pass filters with cut-off wavelength of $1000\ nm$ (FESH1000, Thorlabs), while F5 is a long-pass filter with cut-on wavelength of $900\ nm$ (FELH0900, Thorlabs).

\subsection{Results}
\subsubsection{Optimal Resolution Under Different Numerical Apertures }
According to the Rayleigh criterion, the optimal resolution of an up-conversion imaging system is 
\begin{equation}
\label{equ:ruili}
    {d_R} = 1.22\frac{{\lambda f}}{{{D_{eff}}}}= 0.61\frac{\lambda }{{NA}},
\end{equation}
where $\lambda$ is the illumination wavelength, $f$ is the collection lens focal length, and $D_{eff}$ is the effective aperture. In our up-conversion imaging system with incoherent illumination, the illumination wavelength is fixed at $9.625\ \mu m$ (determined by the center wavelength of the up-conversion). The effective aperture equals the smaller of either the crystal aperture or the pump beam waist in the crystal.

\begin{figure*}
  \centering\includegraphics[width=\textwidth]{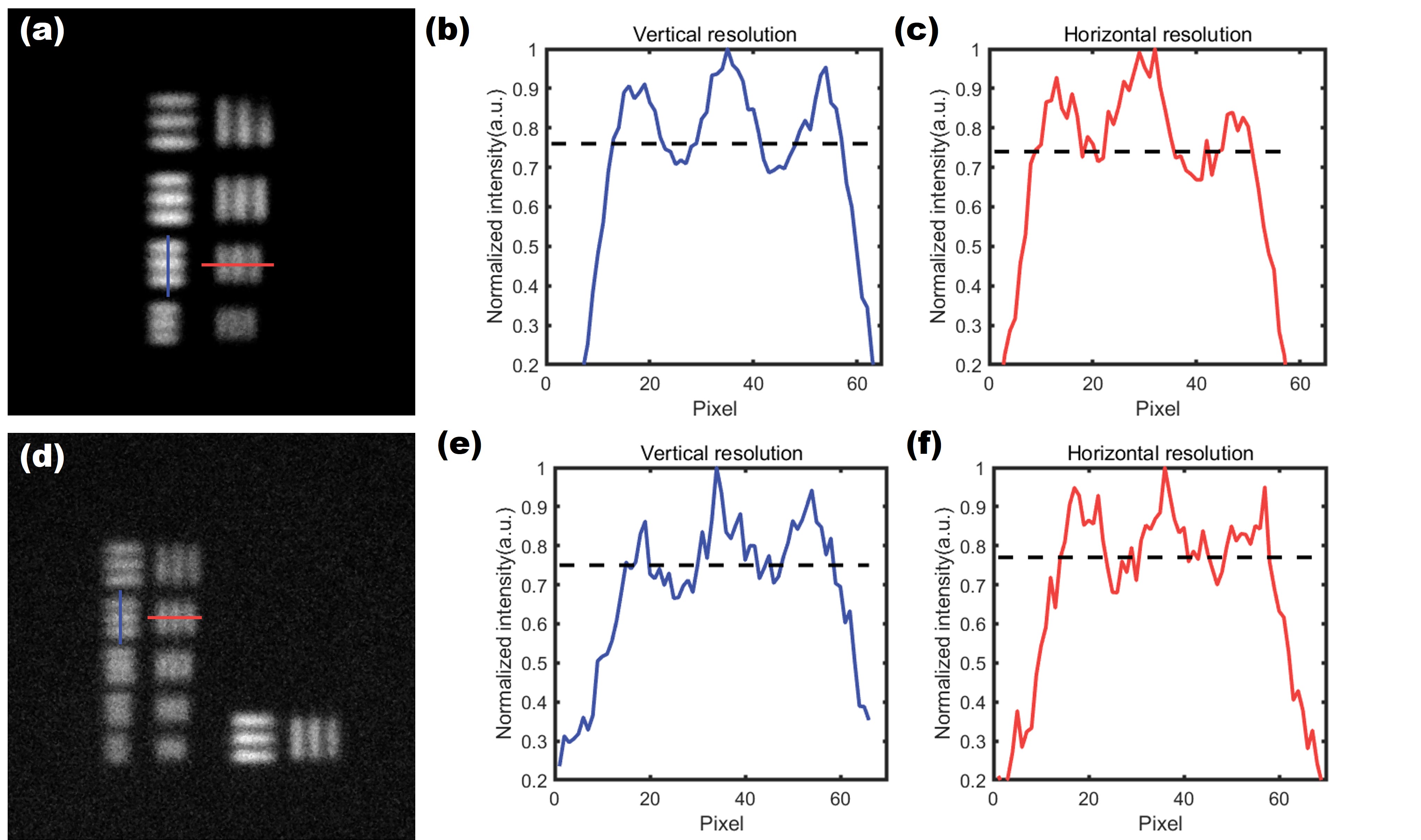}
  \caption{Up-conversion imaging results demonstrating optimal resolution in both lens configurations. (a) Up-converted image of elements 1-1 to 1-4 of USAF-1951 resolution target, using L3 with a focal length of $100\ mm$. (b, c) Normalized intensity distribution along blue and red lines in (a), respectively. (d) Up-converted image of elements 2-1 to 2-6 of USAF-1951 resolution target, using L3 with focal length of $50\ mm$.(e, f) Normalized intensity distribution along blue and red lines in (d), respectively.}
  \label{fig:res}
\end{figure*}

We first investigated the optimal resolution under two different collection lens focal lengths. \textbf{Figure \ref{fig:res}(a)} shows the up-converted image of elements 1-1 to 1-4 in USAF-1951 resolution target acquired by using L3 with a focal length of $100\ mm$. Intensity distribution are acquired by calculating the average pixel values of light ($\overline{I}_{\mathrm{max}}$) and dark ($\overline{I}_{\mathrm{min}}$) lines. When the valley-to-peak ratio $\overline{I}_{\mathrm{min}}/\overline{I}_{\mathrm{max}}$ is less than 0.8, the imaging system is capable of distinguishing the corresponding line pairs according to the Rayleigh criterion. As presented in \textbf{Figure \ref{fig:res}(b, c)}, the intensity of dark lines falls below the black dotted lines($0.8\overline{I}_{\mathrm{max}}$), illustrating that the element 1-3 in the resolution target can be distinguished. Since the image of element 1-4 does not meet the criteria, the resolution of the imaging system equals the line pair width of element 1-3, $396.8\ \mu m$, which is slightly higher than the theoretical resolution limit $391.4\ \mu m$ (please refer to supplementary materials for more details). The up-converted image of elements 2-1 to 2-6 acquired by using L3 with a focal length of $50\ mm$ is shown in \textbf{Figure \ref{fig:res}(d)}. Following the same criterion, the minimum resolvable line pair in the image is element 2-3, with line pair width of $198.4\ \mu m$, which is also close to the theoretical resolution limit $195.7\ \mu m$.

\begin{figure*}
  \centering\includegraphics[width=\textwidth]{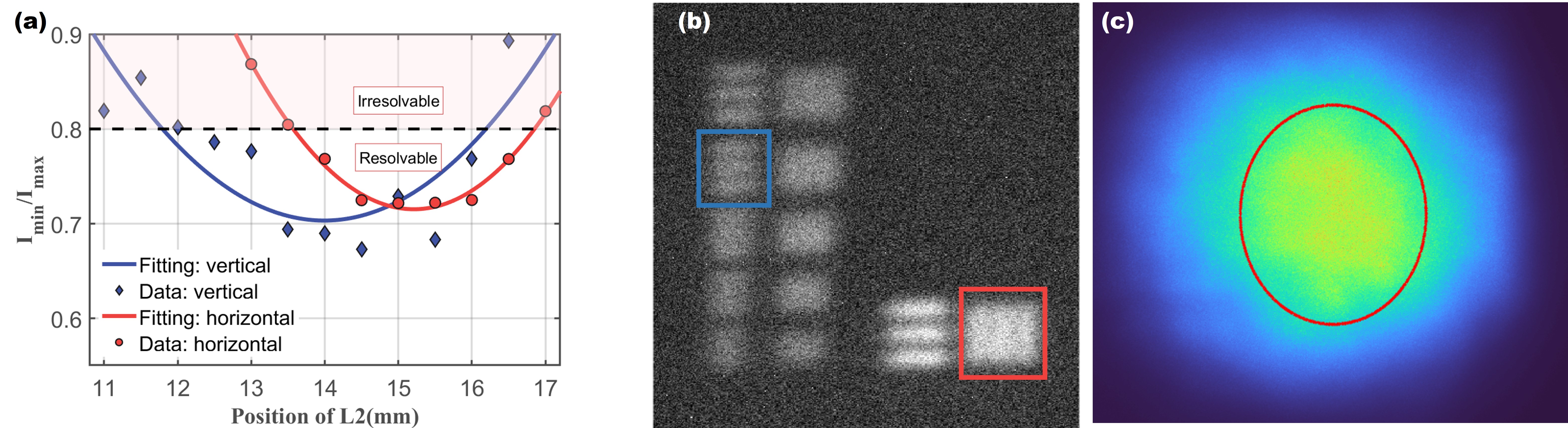}
  \caption{The dependence of imaging resolution on the pump beam waist with L3 focal length of $50\ mm$. (a) The valley-to-peak ratio of element 2-3 in up-conversion images, where blue represents the vertical resolution, red denotes the horizontal resolution, scatter points indicate experimental data, and curves show fitted results. (b) Up-conversion image captured with L2 positioned at $12\  mm$. The line pairs in the blue box and in the red box represent vertical resolution and horizontal resolution, respectively. (c) Intensity profile of the pump beam with L2 positioned at $12\  mm$. The red ellipse is defined by the effective aperture. }
  \label{fig:waist}
\end{figure*}

With the focal length of lens L3 fixed at $50\ mm$, the beam waist of the pump light was adjusted by moving L2, and its beam profile was recorded. The valley-to-peak ratio data of element 2-3 and the fitting curves are shown in \textbf{Figure \ref{fig:waist}(a)}. When L2 was positioned at $12\ mm$, the optimal vertical resolution is determined by element 2–3, while the optimal horizontal resolution is given by element 2–1, as shown in \textbf{Figure \ref{fig:waist}(b)}, yielding the effective aperture is $2.96\ mm$ vertically and $2.35\ mm$ horizontally. \textbf{Figure \ref{fig:waist}(c)} displays the pump beam profile, where the red ellipse represents the effective aperture. For L2 at $13.5\ mm$, the effective aperture is $3.00\ mm$ vertically and $2.96\ mm$ horizontally, while for L2 at $15\ mm$, the effective aperture is $3.00\ mm$ in both directions. At the effective aperture boundary, the average intensity of the pump beam was 0.57 times its peak intensity. This indicates that when the pump beam serves as the effective aperture of the up-conversion imaging system, the beam waist should be more appropriately defined as 0.57 of peak power.

\subsubsection{Depth of Field \& Astigmatism}
The above results demonstrate the system's resolution under optimal conditions. However, in practical applications, some factors can degrade the resolution, preventing it from approaching the theoretical limit. In this system, the primary factors contributing to aberrations are depth of field (DOF) and astigmatism in our opinion. DOF and astigmatism in up-conversion imaging systems can be expressed as (detailed derivation are shown in supplementary materials):
\begin{equation}
{d_{DoF}} =  \frac{{\lambda }}{{N{A^2}}} = \frac{{4\lambda {f^2}}}{{{D^2}}}
\end{equation}
\begin{equation}
{d_{a}} = {M^2}\frac{{{f_1}{{\sin }^2}\theta }}{{{n_1} - 1}} + \frac{{{f_2}{{\sin }^2}\left( {\frac{\theta }{M}} \right)}}{{{n_2} - 1}}
\label{equ:ast}
\end{equation}
Among them, $M$ is the lateral magnification, $f_i$ and $n_i \left( i=3,\ 4 \right)$ represent the focal length and refractive index of lenses L3 and L4, respectively, and $\theta$ is the aperture angle of the imaging target to the lens L3.

\begin{figure*}
  \centering\includegraphics[width=\textwidth]{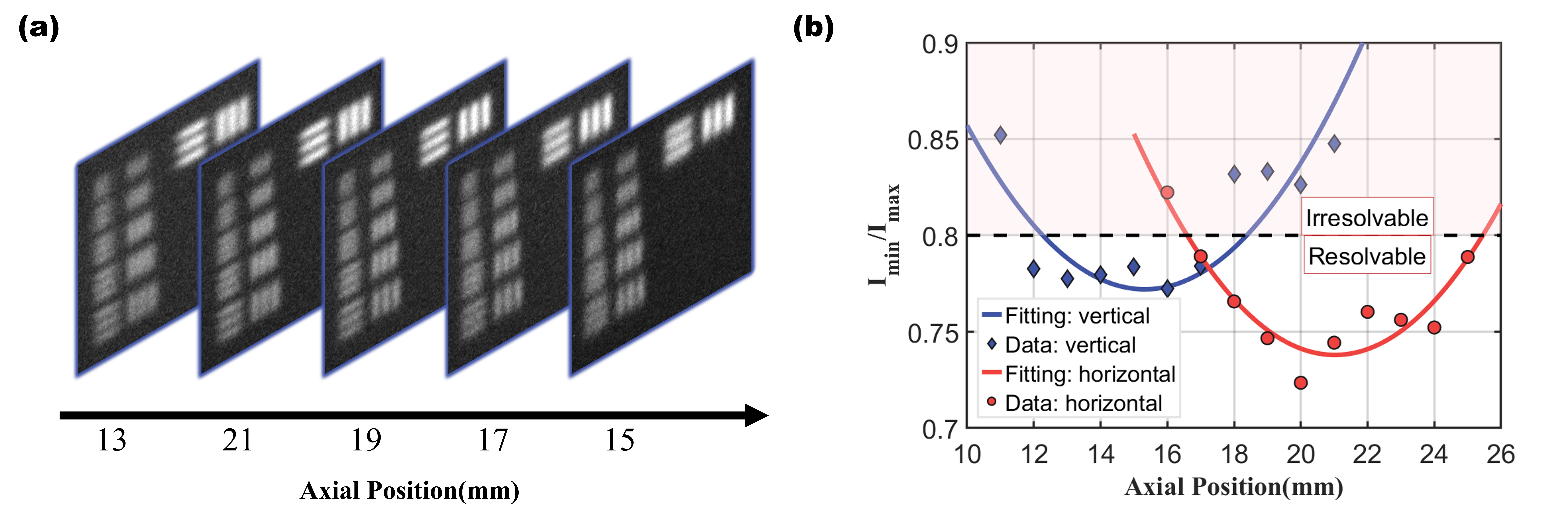}
  \caption{The depth of field and astigmatism of the up-conversion imaging system while the focal length of L3 is $50\ mm$. (a) Up-conversion images of imaging object at different axial positions. For better visualization, these images have been vertically flipped, while the orientation of the line pairs remains unchanged. (b) The valley-to-peak ratio of element 2-3 in up-conversion images, where blue represents the vertical resolution, red denotes the horizontal resolution, scatter points indicate experimental data, and curves show fitted results. }
  \label{fig:dep}
\end{figure*}

Using L3 with a focal length of $50\ mm$, we captured up-conversion images of group 2 in the resolution target at different axial positions, as shown in \textbf{Figure \ref{fig:dep}(a)}, and analyzed the variation in resolution. The data and fitting curves of the image resolution are presented in \textbf{Figure \ref{fig:dep}(b)}, where the diamond-shaped dots and the round dots correspond to the vertical resolution and the horizontal resolution data respectively. The axial deviation of the imaging object from the front focal plane of the 4f system leads to resolution degradation. Fitting curves show that elements 2-3 are vertically resolvable within $6.1\ mm$ and horizontally within $8.9\ mm$ along the optical axis, while theoretical calculations show that the DOF is $10.67\ mm$. The horizontal depth of field shows good agreement with theoretical predictions, while further optimization of the pump beam waist profile to achieve better uniformity could help reduce the discrepancy observed in the vertical direction.The asymmetry in resolution degradation along different directions at the same position is attributed to astigmatism, which is discussed in detail in the supplementary materials. \textbf{Figure \ref{fig:dep}(b)} shows that the optimal resolution positions for the two directions locate at $15.3\ mm$ and $21.0\ mm$ respectively. It is evident that the astigmatism of this case is $5.7\ mm$. The target position has an angle of approximately $4^\circ$ with respect to the center of the lens, substituting it into \textbf{Equation \ref{equ:ast}} yields the theoretical astigmatism is $5.04\ mm$, which is reasonable consistent with our experimental result.

As for the case while the focal length of L3 is $100\ mm$, due to the limit of the displacement table, we only measured the DOF when moving to one side, which is $18.5\ mm$, corresponding to an actual DOF of $37\ mm$. The theoretical calculation result is $42.67\ mm$, which is  basically consistent with our experimental results. The experimental and theoretical results of astigmatism are $21.71\ mm$ and $19.45\ mm$, respectively. The calculation results of the equations we proposed are reasonable consistent with experimental results under different focal lengths of the lens L3. For more information about the $100\ mm$ case, please refer to the supplementary materials.

\section{Discussion}
With the pump wavelength fixed at $1078.22\ nm$, we systematically investigated factors affecting optimal resolution and causing resolution degradation. The minimum resolvable line width of $99.2\ \mu m$ has been achieved. The resolution exhibits proportional scaling with lens focal length predicted by the Rayleigh criterion. This precise agreement demonstrates that other components in our imaging system operate at their theoretical limits. Resolution enhancement remains achievable through reducing the focal length.

The conversion efficiency of high spatial frequency components in traditional definition methods is too low compared to zero frequency components, which can affect resolution. The intensity on the effective apertures boundary is about $57\%$ of the peak intensity fitting by our experimental results. Compared to prior works, where the intensity of the waist boundary is considered as $1/e$ \cite{junaid2019video} or $1/e^2$ \cite{zheng2024mid} of the peak intensity, there are considerable differences. This work provides an experimental basis for designing the pump beam waist.  

We have studied for the first time the effect of aberration on resolution in an UCD system, and our equations can accurately predict the size of DOF and astigmatism. The theoretical calculation results are consistent with the experimental results under two different focal lengths. This work provides a new perspective for the further design and practical application of up-conversion systems.

In summary, we have systematically studied the up-conversion imaging of LWIR incoherent thermal radiation targets for the first time and achieved high-resolution detection. The method proposed in this work for calculating various parameters that affect resolution will provides a new perspective for the further design and practical application of up-conversion systems.

\begin{backmatter}
\bmsection{Funding}
We would like to acknowledge the support from the National Key Research and Development Program of China (2022YFB3903102, 2022YFB3607700), National Natural Science Foundation of China (NSFC)(62435018), Innovation Program for Quantum Science and Technology (2021ZD0301100), USTC Research Funds of the Double First-Class Initiative(YD2030002023), and Research Cooperation Fund of SAST, CASC (SAST2022-075).

\bmsection{Acknowledgment}
Z.-Q.-Z. Han, X.-H. Wang, Z.-Y. Zhou and B.-S. Shi investigated the references and came up with the idea. 
Z.-Q.-Z. Han and X.-H. Wang designed the experiments and were involved in building the experimental optical paths, experiment construction, data acquisition and figure visualization. 
J.-P. Li, B.-W. Liu, Z.-H. Zhou and H. Zhang helped collect some of the data. 
Y.-H. Li designed the electronics system. 
Z.-Y. Zhou and B.-S. Shi supervised the whole work and provided the funds. 
Z.-Q.-Z. Han and X.-H. Wang contribute equally in this work. 
All authors have contributed to writing this article. 
We also thank Dr. Ren-Hui Chen, Yue-Wei Song and Zhi-Cheng Guo for helpful discussion.

\bmsection{Disclosures}
The authors declare no conflicts of interest.

\bmsection{Data Availability}
Data may be obtained from the authors upon reasonable request.

\bmsection{Supplemental document}
A supplemental document must be called out in the back matter so that a link can be included. For example, “See Supplement 1 for supporting content.” Note that the Supplemental Document must also have a callout in the body of the paper.

\end{backmatter}


\bibliography{sample}

\end{document}